\documentclass[11pt]{article}
\usepackage{moriond,epsfig}
\usepackage[latin1]{inputenc}
\usepackage[OT1]{fontenc}

\def\vev#1{\langle#1\rangle}

\def\hpc{$h^{-1}$Mpc }
\def\hpcp{$ h^{-1}$Mpc. }
\def\hpcv{$ h^{-1}$Mpc, }
\def\ie{{\it i.e.\ }}
\def\delg{\delta_g} 

\def\be{\begin{equation}}
\def\ee{\end{equation}}
\def\bea{\begin{eqnarray}}
\def\eea{\end{eqnarray}}

\begin{document}
\vspace*{6.5cm}
\title{TESTING GRAVITY ON LARGE SCALES:  \\ THE SKEWNESS OF THE GALAXY DISTRIBUTION AT Z$\sim$1}

\author{C. MARINONI and the VVDS TEAM\footnote{L. Guzzo, 
A. Cappi,
O. Le F\`evre, 
A. Mazure,
B. Meneux, 
A. Pollo, 
A. Iovino, 
H.J. McCracken, 
R. Scaramella, 
S. de la Torre, 
J. M. Virey, 
D. Bottini, 
B. Garilli, 
V. Le Brun, 
D. Maccagni, 
J.P. Picat, 
M. Scodeggio, 
L. Tresse,  
G. Vettolani, 
A. Zanichelli, 
C. Adami, 
S. Arnouts, 
S. Bardelli, 
M. Bolzonella, 
S. Charlot, 
P. Ciliegi, 
T. Contini, 
S. Foucaud, 
P. Franzetti, 
I. Gavignaud, 
O. Ilbert, 
F. Lamareille, 
B. Marano, 
G. Mathez, 
R. Merighi, 
S. Paltani,
R. Pell\`o, 
L. Pozzetti, 
M. Radovich, 
D. Vergani, 
G. Zamorani, 
E. Zucca,  
U. Abbas, 
M. Bondi, 
A. Bongiorno, 
J. Brinchmann, 
A. Buzzi, 
O. Cucciati, 
L. de Ravel, 
L. Gregorini, 
Y. Mellier, 
P. Merluzzi, 
E. Perez-Montero, 
P. Taxil, 
S. Temporin, 
C.J. Walcher}
}

\address{Centre de Physique Th\'eorique, Universit\'e de Provence, CNRS-Luminy \\
Case 907, Marseille, France} 

\maketitle\abstracts{We study the
  evolution of the low-order moments of the galaxy overdensity distribution
  over the redshift interval 0.7$<$z$<$1.5. We find that the
  variance and the normalized skewness evolve over this
  redshift interval in a way that is remarkably consistent with
  predictions of first- and second-order perturbation theory.  This
  finding confirms the standard gravitational instability paradigm
  over nearly 9 Gyrs of cosmic time and demonstrates the importance of
  accounting for the non-linear component of galaxy biasing to
  avoid disagreement between theory and observations.}

\section{Introduction}

Determining the value of the parameters entering into the  Friedmann-Robertson-Walker
model is a classical problem of cosmology which has recently been addressed
with unprecedented accuracy \cite{dunk}.
A large variety of independent data are suggestive of a new 
physical scenario, rich in philosophical implications: it seems that 
we live in a universe where ordinary baryonic matter is a minority ($\sim$1/6) of all matter, 
where matter itself is a minority ($\sim$1/4) of all energy, where geometry is spatially 
flat and the metric expansion is presently accelerated.
However, to make sense of these measurements, a mysterious dark energy component has been 
added to an already elusive ingredient, \ie dark matter.

Since fixing model parameters is not measuring and, as such, it can hardly give us 
insight into the physical nature of the phenomenon investigated,  it is  critical to 
understand whether what we interpret as new cosmic components is rather the smoking 
gun of the failure of our theoretical model on large cosmological scales
\cite{dva,am,buz}.

\begin{figure*}
\includegraphics[width=83mm,angle=0]{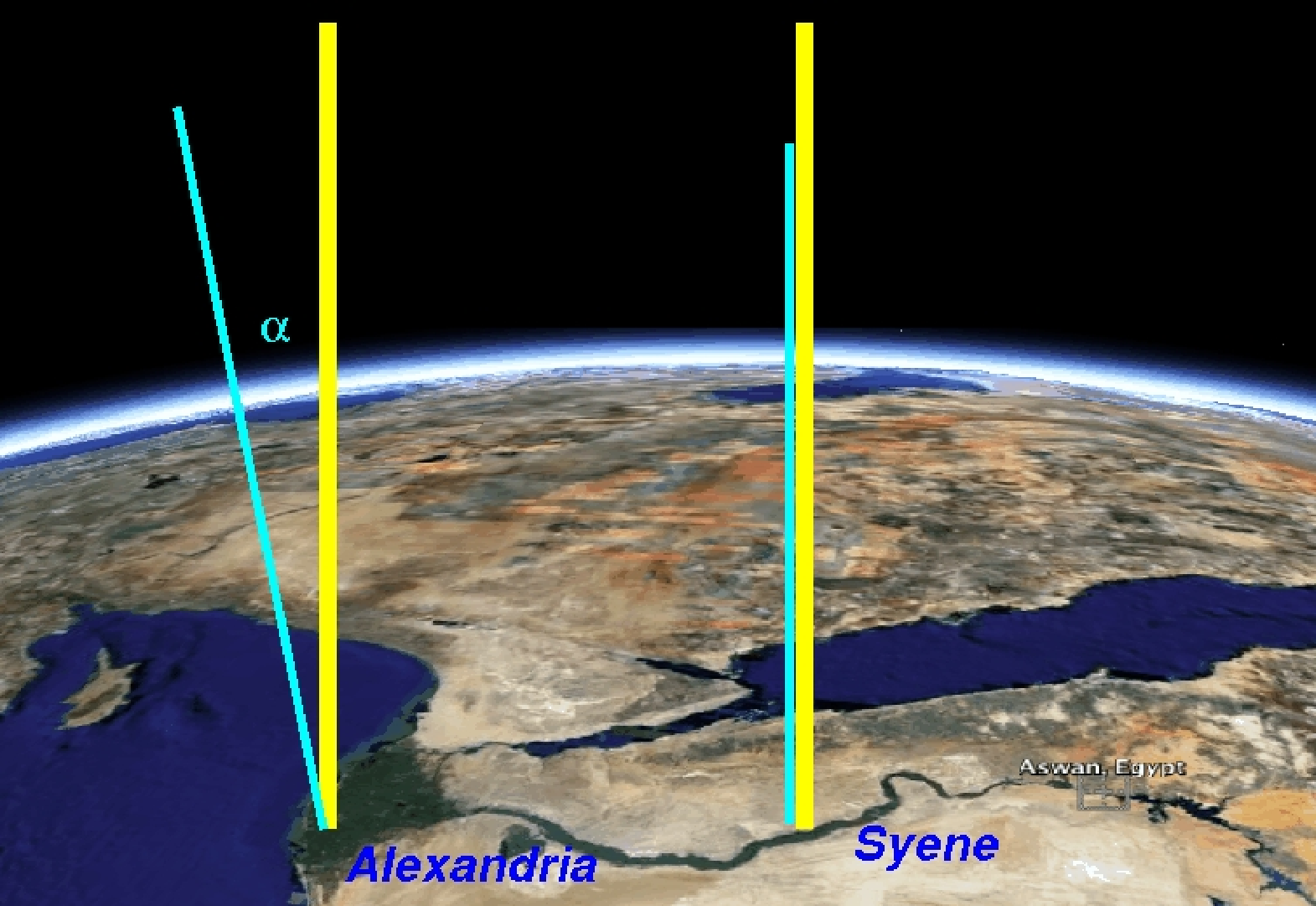}
\includegraphics[width=80mm,angle=0]{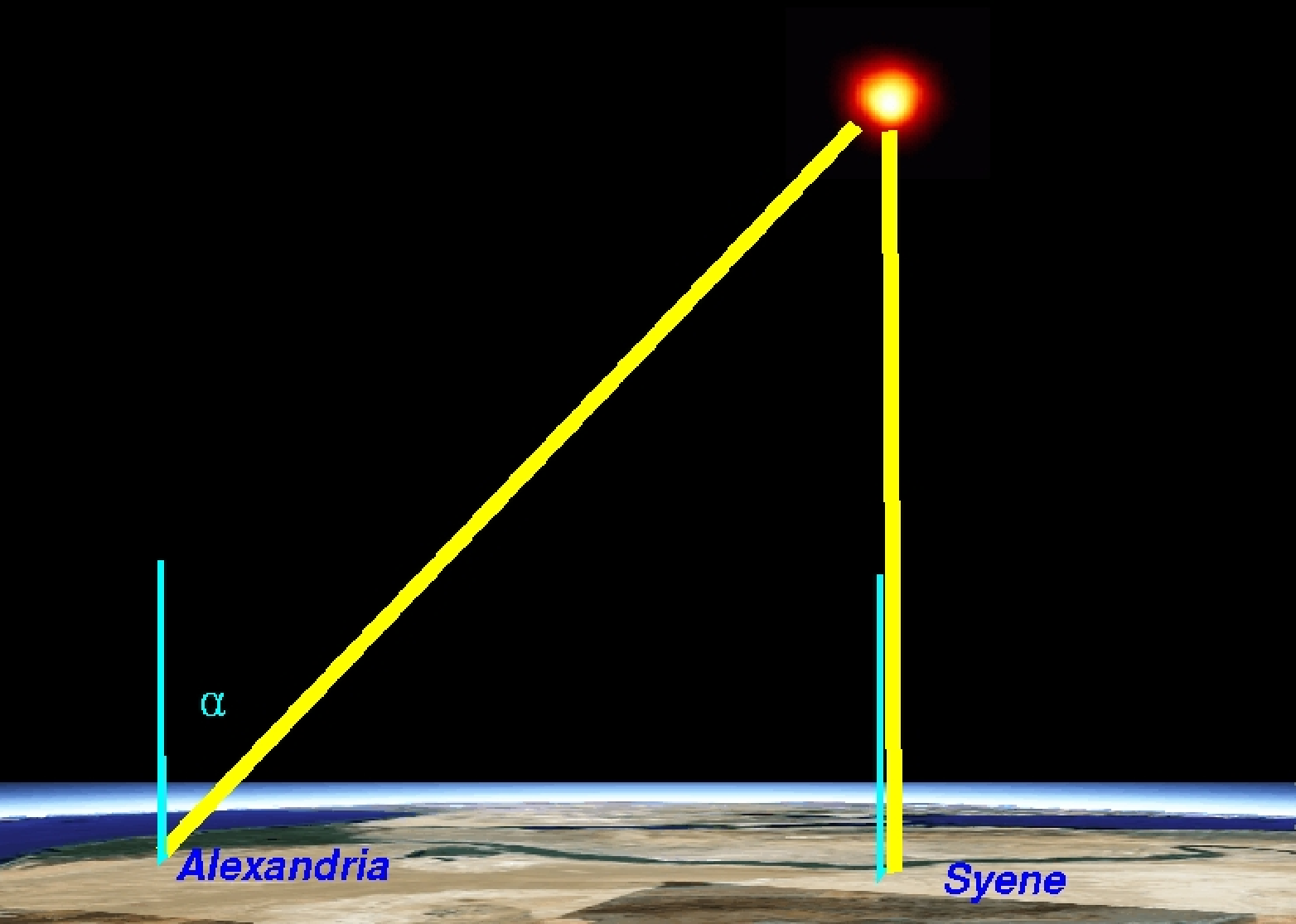}
\caption{World models of Eratosthenes ({\em left}) and Anaxagoras ({\em right}). 
Note that the same astronomical observations were available to the two Greek scholar
{\em i.e.} that at noon, the sun rays were differently inclined  
in Syene and Alexandria (here only pictorially represented) the day of the summer solstice.
But data do not speak by themselves; rather they must be interpreted 
within a theoretical framework. 
Without an independent test of the hypothesis that the 
earth is spherical,  Eratosthenes' is not a physical measure but a 
model-dependent  parameter estimation.}
\label{fig0}
\end{figure*}

An example will better illustrate how important is testing
the hypotheses underlying the standard
model of cosmology as well as the soundness of the assumptions
 implicit in cosmological 
``measurements". Eratosthenes is remembered for a 
technique he introduced which enabled him to compute the first reliable determination 
of the radius of the earth. He interpreted available data (the different inclination angle 
of sun rays at noon in Syene and Alexandria the day of summer solstice) assuming 
that the earth {\em is} spherical, that the two towns are on the same meridian and 
that the sun is far enough that its rays are almost perfectly parallel (see Fig. \ref{fig0}). 
On the basis
of these hypotheses he estimated the radius of the earth with a
precision greater than the accuracy currently attained in measuring dark energy. 
What is not often emphasized is that these  same `high quality' data were available to 
another Greek scholar, Anaxagoras who lived nearly two centuries before. 
By interpreting them  assuming that the earth {\it is} flat and that the different 
inclination of the rays is due to 
the sun proximity (see right panel of Fig \ref{fig0}), he concluded, with 
spectacular precision, that the sun is as big as the Peloponnese.

The picture in which gravity, as described by general relativity, is
the engine driving cosmic growth is generally referred to as the
gravitational instability paradigm (GIP). However plausible it may
seem, it is critical  to test its validity. In the local universe the
GIP paradigm has been shown to make sense of a vast amount of
independent observations on different spatial scales from galaxies to
superclusters of galaxies \cite{pea01,teg06}.  Deep redshift
surveys now allow us to test whether the predictions of this
assumption are also valid at earlier epochs \cite{mar08}.

We test the role of gravity in shaping density
inhomogeneities by using three-dimensional maps of the distribution of
visible matter revealed by the VIMOS-VLT Deep Survey \cite{lef05,mar08} over the large
redshift baseline $0<z<1.5$ (see Massey et al. \cite{mas07} for three
dimensional cartography of mass overdensities in the COSMOS field). 

We explore the mechanisms governing this growth by comparing the time
evolution of the low-order moments of the galaxy PDF, ({\em i.e.}  the
{\it variance} amplitude $<\delta_g^2>$ and the {\it normalized
  skewness} $S_3=<\delta_g^3>_c/<\delta_g^2>^{2}$) with the
corresponding quantity theoretically predicted for matter fluctuations
in the linear and semi-linear perturbation regime.
This provides a test of GIP-specific predictions at as-yet unexplored epochs
that are intermediate between the present era and the time of
decoupling.  Knowledge of the precise growth history of density
inhomogeneities provides also a way to test the theory of
gravitation~\cite{lin05}.

In addition to the statistical approach presented in this paper, we
have recently addressed this same issue also from a dynamical point of
view. We have used linear redshift-space distortions in the VVDS-{\it
Wide} data to measure the growth rate of matter fluctuations at
$z\sim 0.8$ \cite{nature07}.  This approach offers promising prospects
for determining the cause of cosmic acceleration in the near future
\cite{lin07}.

\section{A cosmographical tour up to $z=1.5$}

By using the VVDS data we have reconstructed, for the first time,  
the three-dimensional map of large-scale galaxy fluctuations  to $z=1.5$.  
The $I \leq 24$ sample is
characterized by an effective mean inter-particle separation of
($\vev{r} \sim 5.1$ \hpc) in the redshift range 0$<z<$1.5]. For
comparison, this sampling is better (denser) than the early CfA1 survey
($\vev{r}\sim 5.5$$ h^{-1}$Mpc) used by Davis \& Huchra \cite{dav81} to reconstruct
the 3D density field of the local Universe (\ie out to $\sim$ 80 \hpc).
Also, at the median depth of the VVDS survey, \ie in the redshift interval
$0.7<z<0.8$, the mean inter-particle separation is 4.4 \hpcv a value
nearly equal to the 2dFGRS at its median depth.

\begin{figure*}
\begin{center}
\includegraphics[width=160mm,angle=0]{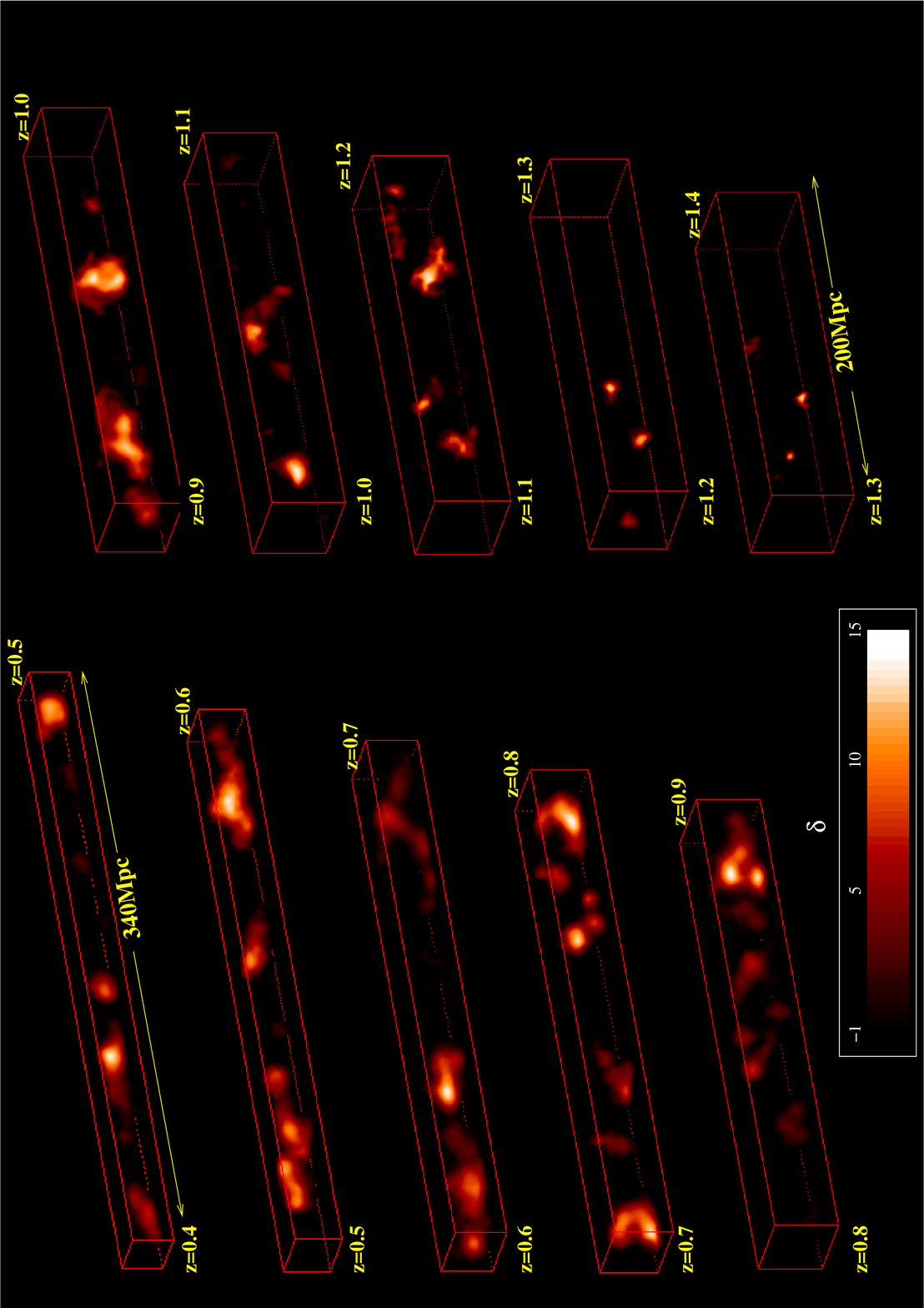}
\caption{ The reconstructed density field for $0.4<z<1.4$, as traced by
  the galaxy distribution in the VVDS-Deep redshift survey to $I \leq
  24$.  This figure preserves the correct aspect ratio between
  transverse and radial dimensions.  The mean inter-galaxy separation
  of this sample at the typical depth of the VVDS ($z=0.75$) is
  $4.6$\hpcv comparable to local redshift surveys as the 2dFGRS.  The
  galaxy density distribution has been smoothed using a 3D Gaussian
  window of radius $R=2$\hpc and noise has been filtered away using a
  Wiener filtering technique.  
  Only fluctuations above a signal-to-noise threshold of
  $2$ are shown.  The accuracy and robustness of the reconstruction
  methods have been tested using realistic mock catalogues (Pollo et al. 2005, 
Marinoni et al. 2005).}
\end{center}
\label{fig1}
\end{figure*}

The recovered galaxy overdensity field is presented in Fig. \ref{fig1}.
Fluctuations have been smoothed on a scale $R=2$\hpcp  Only density
contrasts with signal-to-noise ratio $S/N>2$ are shown.

\begin{figure*}
\begin{center}
\includegraphics[width=110mm,angle=-90]{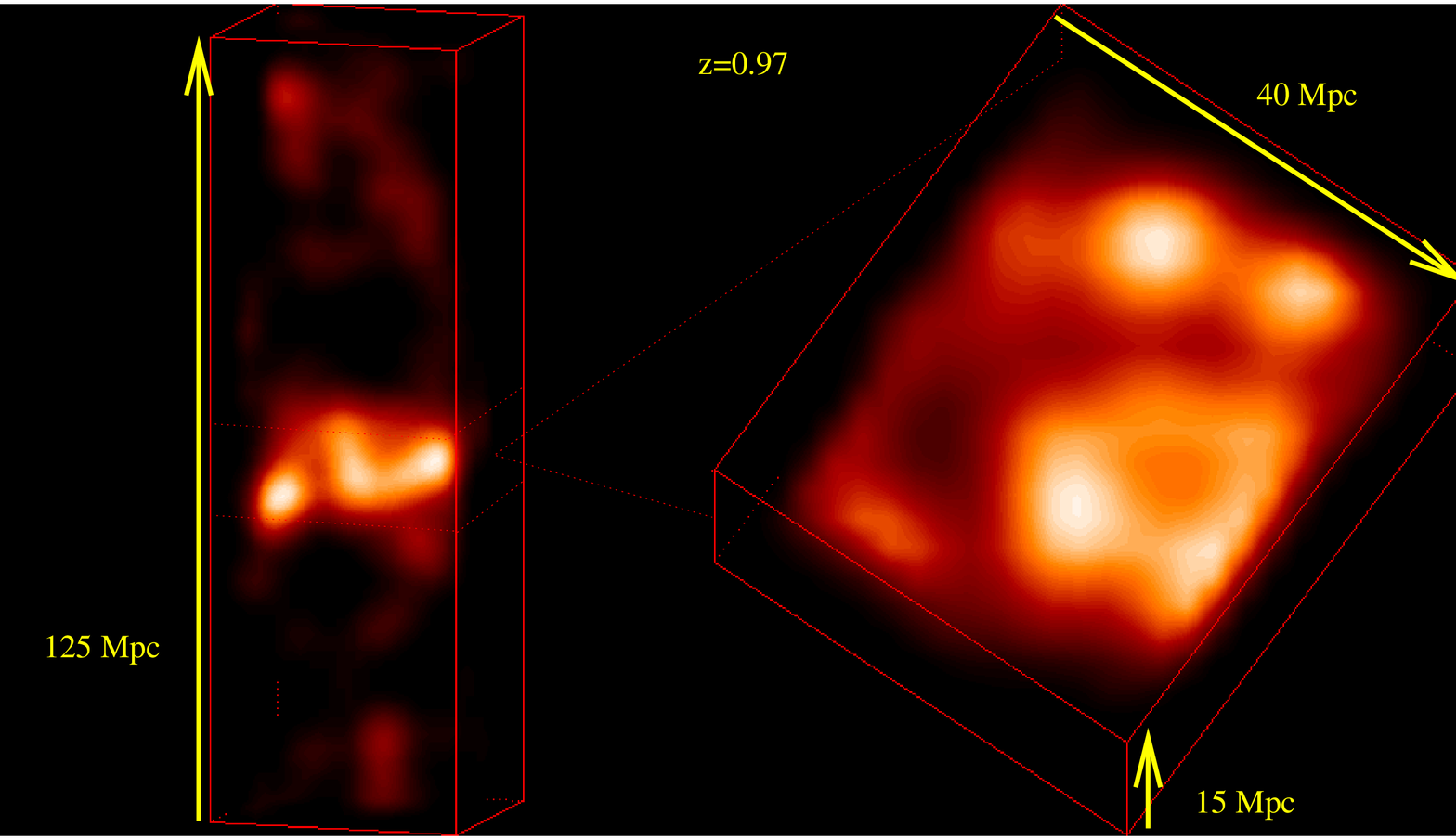}
\caption{
Density distribution and properties of a large-scale planar structure 
at $z=0.97$, that completely fills the VVDS-02h field-of-view.
}
\label{fig2}
\end{center}
\end{figure*}

A remarkable feature of this {``\em geographical''} exploration of the
Universe at early cosmic epochs is the abundance of large-scale
structures similar in density contrast and size (at least in one
direction) to those observed by local surveys.  In particular, it is
tempting to identify qualitatively a few filament-like density
enhancements bridging more condensed structures along the line of
sight, although the survey transverse size is still too small to fully
sample their extent.  Nevertheless, it is interesting to notice that
these apparently one-dimensional structures remain coherent over scales
$\sim 100$\hpcv separating low-density regions of similar size. Figs.
\ref{fig1} and \ref{fig2} visually confirm that the familiar web pattern observed in
the local Universe is not a present-day transient phase of the galaxy
spatial organization but it is already well-defined at $\sim 1.5$ when
the Universe was $\sim 30\%$ its present age.  
This implies that large-scale features of the galaxy
distribution essentially reflects the long-wavelength modes of the
initial power spectrum, in agreement with theoretical predictions of
the CDM hierarchical scenario.  Numerical simulations of large scale
structure formation in fact show that the present-day web of filaments
and walls is actually present when the universe was in embryonic form
in the overdensity pattern of the initial fluctuations, with subsequent
linear and non-linear gravitational dynamics just sharpening its
features \cite{bon96} \cite{spri05}.

The limited angular size of the survey is exemplified by a dense
``wall'' at $z=0.97$ that stretches across the whole survey solid angle
($0.7 \times 0.7$ deg) (see Fig. \ref{fig2}). This two-dimensional structure is
coherent over more than $\sim 30$\hpc (comoving) in the transverse
direction, is only $\sim 10$\hpc thick along the line of sight, and has
a mean overdensity $\delta_g=2.4 \pm 0.3$.  This makes it similar to
the largest and rarest structures observed in the local Universe, such
as the Shapley concentration \cite{scar89}. By applying
a Voronoy-Delaunay cluster finding code \cite{mardav02}, we find 10
distinct groups in this structure, with between 5 and 12 galaxy members
each (down to the limiting magnitude I=24), for a total of 164
galaxies.  If one considers the evolution of {\it mass} fluctuations in
the standard $\Lambda$CDM model, the probability of finding a structure
with similar {\it mass} overdensity at such early times ($0.9<z<1$)
would be nearly 4 times smaller than today: one such {\it mass
  fluctuation} would be expected in a volume of $\sim 3\cdot10^{6}
h^{-3}$Mpc$^3$, \ie nearly 5 times larger than our surveyed volume up
to $z \sim 1$.  In fact, as shown by Marinoni et al 2005 \cite{mar05} finding
such a {\it galaxy} overdensity is not so unusual: it is clear evidence
that the {\it biasing} between galaxies and matter at these epochs is
higher than today. This makes fluctuations in the galaxy distribution
to be highly enhanced with respect to those in the mass.

\section{Testing gravitational instability 
with the low-order moments of the PDF}

We have used the density maps presented in Fig. \ref{fig1}. to reconstruct the
Probability Distribution Function of galaxy fluctuations on large cosmological 
scales \cite{mar08} and to study the  evolution of its low-order statistical moments, 
\ie variance and skewness.

To facilitate comparison between local and high redshift results 
we estimate these quantities for a volume-limited sample of VVDS 
galaxies with $M_{B} \leq -20 +5 \log h$ (i.e. for a sample of test 
particles with median luminosity $\sim 2L*$). Moreover,
since in perturbation theory higher order cumulants are predicted to
be a function of the variance, we will always consider in the following 
the normalized skewness  $S_3=\vev{\delg^3}_c/\sigma^4$.

\begin{center}
\begin{figure}
\begin{center}
\includegraphics[width=120mm,angle=0]{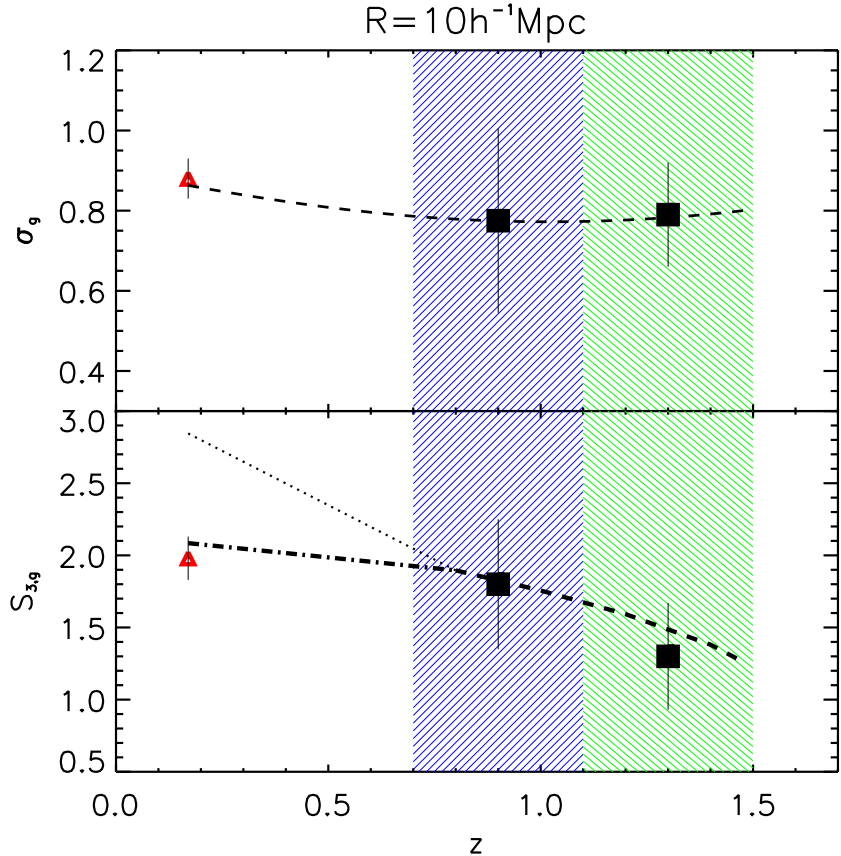}
\caption{Evolution of the {\it r.m.s} deviation (top) and skewness
  (bottom) of the PDF of galaxy fluctuations on a scale $R=10$\hpcp The
  filled squares correspond to two volume-limited samples from the VVDS
  with $M_{B}< -20+5\log h$ covering the redshift intervals indicated
  by the shaded regions.  Triangles correspond to the 2dFGRS
  measurements at $z\simeq 0.15$ (Croton et al. 2005), from a sample including
  similarly bright galaxies.  Error bars give 68\% confidence errors,
  and, in the case of VVDS measurements, include the contribution from
  cosmic variance. The dashed lines in both panels show the theoretical
  predictions for the evolution of the variance (Eq. \ref{var1}) and
  skewness (Eq. \ref{skew2}) inferred using VVDS measurement of
  biasing (Marinoni et al. 2005).  Predictions for the skewness (based on the
  $(b_1(z),b_2(z))$ measurements in the redshift range $0.7<z<1.5$
  have been extrapolated to $z\sim 0$
  using the local (2dFGRS) biasing measurements of Verde et al. 2002
  (linear bias, dotted line) and of Gazta\~{n}aga et al. 2005
  (quadratic bias with $b_2/b_1=-0.34$, dot-dashed line).}
\label{fig5}
\end{center}
\end{figure}
\end{center}

Fig. \ref{fig5} shows the evolution of the {\it rms} fluctuation
and the normalized skewness on a scale $R=10$\hpcv as measured from the
VVDS volume-limited sub-samples.  Errors have been computed using the
50 fully-realistic mock catalogs of VVDS-Deep discussed in Pollo et al.
(2005).  This allows us to include an estimate of the contribution of
cosmic variance, which represents the most significant term in our
error budget.

The top panel of Fig. \ref{fig5} shows that the square-root of the
variance, which measures the r.m.s. amplitude of fluctuations in
galaxy counts, is with good approximation constant over the full
redshift baseline investigated: in redshift space, the mean value of
$\sigma_g$ for our volume-limited galaxy samples 
is $0.78\pm 0.09$ for $0.7<z<1.5$.
A similar, nearly constant value is also consistent with the
  value estimated at $z\sim 0.15$ from the 2dF galaxy redshift survey
 \cite{cro04} that is also reported in same figure. 
  This means that over nearly 2/3 of the age of the Universe the
observed fluctuations in the galaxy distribution look almost as
frozen, despite the underlying gravitational growth of mass
fluctuations. This quantifies the visual impression we had from Fig. \ref{fig1},
  that the distribution of galaxies is as inhomogeneous
  at $z\sim 1$ as it is today. 
 
The third moment, which measures asymmetries between under- and
over-dense regions, indicates that the galaxy density field was
non-Gaussian on large scales (10 \hpc) even at these remote epochs
($\sim 4 \sigma$ detection).  In particular we find indication for an
increase of the normalized skewness with cosmic time, when comparing
the VVDS values to the local measurement by 2dFGRS.

\section{Comparison with Theoretical Expectations }

Marinoni et al. \cite{mar05} used the same VVDS sample of luminous galaxies to  measure
the cosmological biasing between matter and galaxy distributions \cite{mar05}.
The key result from that analysis was  that galaxy biasing is non-linear on scales $R=10$\hpcp
and increasing with redshift.

Using this ingredient we can now contrast the 
observed redshift  scaling of the low-order statistical moments of the galaxy PDF
against the theoretical predictions for the evolution of the variance and skewness
of the matter density field. Our goal is to test the consistency 
of some general predictions of the GIP.

Using linear perturbation theory, the scaling of the {\it rms} of galaxy density
fluctuations is

\begin{equation}
\sigma_{g}(z) \sim b_{1}(z)D(z)p(z)\sigma(0) \,\,\, ,
\label{var1}
\label{mia}
\end{equation}
where $b_1$ is the linear term of the biasing function \cite{mar08},
$D(z)$ is the linear growth factor of density
fluctuations, $p(z)$ is the redshift-dependent
Kaiser correction which takes into account the average contribution of
the linear redshift distortions induced by peculiar velocities
\cite{kai87}, and $\sigma(0)$ is the present-day {\it rms} of the 
mass density fluctuations.

In a Universe in which primordial density fluctuations were Gaussian,
the non-linear nature of gravitational dynamics leads to the emergence
of a non-trivial skewness of the local density PDF.  
According to predictions of the non-linear, second-order perturbation theory,
the skewness of the mass distribution is approximately independent of
time, scale, density, or geometry of the cosmological model. Assuming
that its evolution only depends on the hypothesis that the initial
fluctuations are small and quasi-Gaussian and that they grow via
gravitational clustering one derives that, in redshift-distorted space
\cite{hiv95}

\begin{equation}
S_{3}\sim \frac{35.2}{7}-1.15(n+3)
\end{equation}

\noindent where $n$ is the effective slope of the power spectrum on the
scales of interest (i.e. in our case, since $R=10$\hpc, n is
approximately given by -1.2).  Substituting the
relevant expansion terms  of the biasing function, the evolution of the
observed skewness is given by Fry \& Gazta\~{n}aga 1993 \cite{fg93}

\begin{equation}
S_{3,g}\sim b_{1}(z)^{-1} \Big[S_3 + 3 \frac{b_2(z)}{b_1(z)} \Big]. 
\label{skew2}
\end{equation}

The curves in both panels of Fig. \ref{fig5} show that equations (\ref{var1})
and (\ref{skew2}) reproduce extremely well the evolution of variance
and skewness observed within the VVDS. 
Concerning the local measurements from
2dFGRS, the predicted scaling for the skewness continues to show very good
agreement if, even locally, biasing is non-linear as we measured at high redshift
($
\left\langle \frac{b_2}{b_1} \right\rangle=-0.19 \pm 0.04 )$
over the redshift range $0.7<z<1.5$)
and as confirmed by the analysis of  Gazta\~{n}aga et
al.\cite{gaz05} of the 2dFGRS sample ($b_2/b_1=-0.34$)
These results provide an indication of the
consistency, at $z=1$, of some constitutive elements of the standard
picture of gravitational instability from Gaussian initial conditions.
The value of $S_{3,g}$, however,
cannot be consistent with GIP predictions if in the local
universe the simple linear biasing measurement of Verde et
al. \cite{ver02} (\ie  $b_2=0$)   is adopted. 

The results we have  presented  provide the first direct evidence
at $z\sim1$ for the consistency of the GIP hypothesis as described in the
framework of general relativity.  The standard theory of structure
formation via gravitational instability successfully explains the
present day statistics (e.g. Tegmark et al. 2006) and dynamics (e.g.
Peacock et al. 2001) of large scale structures.
We have shown that
observations are fully consistent with these predictions over the
entire redshift baseline $0<z<1.5$  only if the small ($10\%$) yet  crucial  
non-linearities measured in the  biasing relation are taken into account.

\end{document}